\documentclass[aip,pop,reprint,twocolumn,showpacs,floatfix,preprintnumbers,amsmath,amssymb,nofootinbib]{revtex4-1}

\usepackage{graphicx}
\usepackage{dcolumn}
\usepackage{bm}
\usepackage{wasysym}
\usepackage{xfrac}
\usepackage{nicefrac}
\usepackage{courier}
\usepackage{dcolumn}
\usepackage{bigints}
\usepackage{gensymb} 
\usepackage{comment}
\usepackage{hyperref}
\usepackage{fancyhdr}

\makeatletter
\def\frontmatter@makefnmark{%
 \@textsuperscript{%
  \normalfont\@thefnmark
 }%
}%
\makeatother

\begin{document}


\title{Exploring magnetized liner inertial fusion with a semi-analytic model$^\dagger$}
\fancyhf{}
\rhead{Page \thepage\ of 13}
\cfoot{$^\dagger$Journal Reference: Phys. Plasmas {\bf 23}, 012705 (2016); \url{http://dx.doi.org/10.1063/1.4939479}.}
\renewcommand{\headrulewidth}{0.0pt}

\author{R.~D.~McBride}
\affiliation{Sandia National Laboratories, Albuquerque, New Mexico 87185, USA}
\author{S.~A.~Slutz}
\affiliation{Sandia National Laboratories, Albuquerque, New Mexico 87185, USA}

\author{R.~A.~Vesey}
\affiliation{Sandia National Laboratories, Albuquerque, New Mexico 87185, USA}
\author{M.~R.~Gomez}
\affiliation{Sandia National Laboratories, Albuquerque, New Mexico 87185, USA}
\author{A.~B.~Sefkow}
\affiliation{Sandia National Laboratories, Albuquerque, New Mexico 87185, USA}
\author{S.~B.~Hansen}
\affiliation{Sandia National Laboratories, Albuquerque, New Mexico 87185, USA}
\author{P.~F.~Knapp}
\affiliation{Sandia National Laboratories, Albuquerque, New Mexico 87185, USA}
\author{P.~F.~Schmit}
\affiliation{Sandia National Laboratories, Albuquerque, New Mexico 87185, USA}
\author{M.~Geissel}
\affiliation{Sandia National Laboratories, Albuquerque, New Mexico 87185, USA}
\author{A.~J.~Harvey-Thompson}
\affiliation{Sandia National Laboratories, Albuquerque, New Mexico 87185, USA}
\author{C.~A.~Jennings}
\affiliation{Sandia National Laboratories, Albuquerque, New Mexico 87185, USA}
\author{E.~C.~Harding}
\affiliation{Sandia National Laboratories, Albuquerque, New Mexico 87185, USA}
\author{T.~J.~Awe}
\affiliation{Sandia National Laboratories, Albuquerque, New Mexico 87185, USA}
\author{D.~C.~Rovang}
\affiliation{Sandia National Laboratories, Albuquerque, New Mexico 87185, USA}
\author{K.~D.~Hahn}
\affiliation{Sandia National Laboratories, Albuquerque, New Mexico 87185, USA}
\author{M.~R.~Martin}
\affiliation{Sandia National Laboratories, Albuquerque, New Mexico 87185, USA}
\author{K.~R.~Cochrane}
\affiliation{Sandia National Laboratories, Albuquerque, New Mexico 87185, USA}
\author{K.~J.~Peterson}
\affiliation{Sandia National Laboratories, Albuquerque, New Mexico 87185, USA}
\author{G.~A.~Rochau}
\affiliation{Sandia National Laboratories, Albuquerque, New Mexico 87185, USA}
\author{J.~L.~Porter}
\affiliation{Sandia National Laboratories, Albuquerque, New Mexico 87185, USA}
\author{W.~A.~Stygar}
\affiliation{Sandia National Laboratories, Albuquerque, New Mexico 87185, USA}
\author{E.~M.~Campbell}
\affiliation{\mbox{Laboratory for Laser Energetics, University of Rochester, Rochester, New York 14623, USA}}
\author{C.~W.~Nakhleh}
\affiliation{Los Alamos National Laboratory, Los Alamos, New Mexico 87545, USA}
\author{M.~C.~Herrmann}
\affiliation{Lawrence Livermore National Laboratory, Livermore, California 94551, USA}
\author{M.~E.~Cuneo}
\affiliation{Sandia National Laboratories, Albuquerque, New Mexico 87185, USA}
\author{D.~B.~Sinars}
\affiliation{Sandia National Laboratories, Albuquerque, New Mexico 87185, USA}


\date{\today}

\begin{abstract}
In this paper, we explore magnetized liner inertial fusion (MagLIF) [S.~A.~Slutz {\it et al}., Phys. Plasmas {\bf 17}, 056303 (2010)] using a semi-analytic model [R.~D.~McBride and S.~A.~Slutz, Phys. Plasmas {\bf 22}, 052708 (2015)].  Specifically, we present simulation results from this model that: (a)~illustrate the parameter space, energetics, and overall system efficiencies of MagLIF; (b)~demonstrate the dependence of radiative loss rates on the radial fraction of the fuel that is preheated; (c)~explore some of the recent experimental results of the MagLIF program at Sandia National Laboratories [M.~R.~Gomez {\it et al}., Phys. Rev. Lett. {\bf 113}, 155003 (2014)]; (d) highlight the experimental challenges presently facing the MagLIF program; and (e) demonstrate how increases to the preheat energy, fuel density, axial magnetic field, and drive current could affect future MagLIF performance.
\end{abstract}



\maketitle
\thispagestyle{fancy}

\section{\label{sec:intro}Introduction}

The Magnetized Liner Inertial Fusion (MagLIF) concept\cite{Slutz_PoP_2010,Slutz_PRL_2012} is presently being investigated experimentally\cite{Sinars_PRL_2010,Sinars_MRT_PoP_2010,Cuneo_IEEE-TPS_2012,Martin_PoP_2012,McBride_PRL_2012,McBride_PoP_2013,Dolan_RSI_2013,Peterson_PoP_2012,Peterson_PoP_2013,Peterson_PRL_2014,Awe_PRL_2013,Awe_PoP_2014,Sefkow_PoP_2014,Gomez_PRL_2014,Schmit_PRL_2014,Gomez_PoP_2015,Hansen_PoP_2015,Knapp_PoP_2015} using the Z facility\cite{Rose_PRSTAB_2010_EM_model,Savage_PPC_2011} at Sandia National Laboratories. MagLIF is part of a broader class of concepts referred to collectively as magneto-inertial fusion (MIF).\cite{Khariton_UFN_1976,*Khariton_SPU_1976,Mokhov_SPD_1979,Sweeney_NF_1981,Lindemuth_PoF_1981,Lindemuth_NF_1983,Jones_NF_1986,Hasegawa_PRL_1986,Lindemuth_PRL_1995,Kirkpatrick_FT_1995,Degnan_PRL_1999,Siemon_CPPCF_1999,Basko_NF_2000,Kemp_NF_2001,Ryutov_CPPCF_2001,Kemp_NF_2003,Intrator_IEEE-TPS_2004,Garanin_IEEE-MegaGaussConf_2006}  These concepts seek to significantly reduce the implosion velocity and pressure requirements of traditional inertial confinement fusion (ICF)\cite{Nuckolls_Nature_1972,Lindl_PoP_1995,Perkins_PRL_2009,Meezan_PoP_2010,Hurricane_Nature_2014} by using a magnetic field to thermally insulate hot fuel\cite{Landshoff_PR_1949} from a cold pusher and to increase fusion product confinement.

The MagLIF concept at Sandia uses the electromagnetic pulse supplied by the Z accelerator to radially implode an initially solid cylindrical metal tube (liner) filled with preheated (100--300 eV) and premagnetized (10--30~T) fusion fuel (deuterium or deuterium-tritium). The implosion is a result of the fast z-pinch process, where a large gradient in the applied magnetic field pressure operates near the liner's outer surface.\cite{Ryutov_FZP_RMP_2000,Cuneo_IEEE-TPS_2012}  The fuel preheating is accomplished using the multi-kJ, 1-TW, frequency-doubled (527-nm) Nd:glass Z beamlet laser (ZBL),\cite{Rambo_AO_2005,Gomez_PRL_2014} and the fuel premagnetization is accomplished using the Applied {\it B} on Z (ABZ) axial field coil system.\cite{Rovang_RSI_2014}  One- and two-dimensional simulations of MagLIF using the LASNEX radiation magnetohydrodynamics code \cite{Zimmerman_CPPCF_1975} predict that if sufficient liner integrity can be maintained throughout the implosion, then significant fusion yield ($>$100 kJ) can be attained on the Z accelerator when deuterium-tritium fuel is used and the accelerator's Marx generators are charged to 95 kV to obtain a peak drive current of about 27 MA.\cite{Slutz_PoP_2010,Cuneo_IEEE-TPS_2012}

\begin{figure}
\includegraphics[width=0.723\columnwidth]{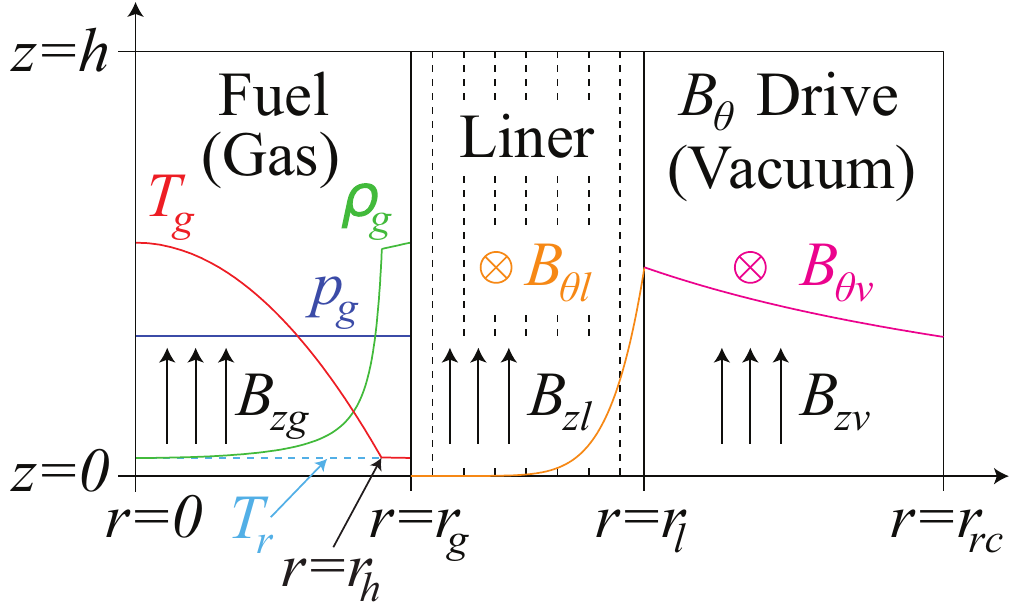}
\caption{\label{fig:schematic} Schematic overview of the semi-analytic MagLIF model. There are three primary regions: a fuel region, a liner region, and a vacuum region.  The system height is $h$; the thermally insulating axial magnetic field, which is initially distributed uniformly over all regions, is $B_{z}$; the radius of the fuel-liner interface is $r_g$; the liner's outer radius is $r_l$; the return current radius is $r_{rc}$; and the azimuthal magnetic field, which drives the cylindrical implosion, is $B_{\theta}$.  Normalized profiles are shown for $B_{\theta}$ in the vacuum region, $B_{\theta v}$ (magenta), and in the liner region,  $B_{\theta l}$ (orange); their analytic expressions are given by Eqs.~3 and 4 in Ref.~\onlinecite{McBride_PoP_2015} ($B_{\theta}$ is assumed to be zero in the fuel region).  The liner region is further divided into multiple concentric liner shells; this discretization is necessary to avoid overdriving the fuel.  Within the fuel region, normalized profiles are shown for the gas pressure, $p_g$ (blue), the gas temperature, $T_g$ (red), the gas density, $\rho_g$ (green), and the radiation temperature, $T_r$ (cyan). The pressure profile is flat throughout the fuel (i.e., we have made an isobaric assumption due to the subsonic nature of MagLIF implosions). The gas temperature and density profiles thus have an inverse dependence to one another; their analytic expressions are given by Eqs.~105--109 in Ref.~\onlinecite{McBride_PoP_2015}.  The radiation temperature is nearly constant across the fuel region.  The fuel region is further divided into a hot spot region from $r=0$ to $r_h$ and a cold dense shelf region from $r_h$ to $r_g$. The gas temperature in the shelf region is equal to the radiation temperature (i.e., the fuel material and the radiation field are assumed to be in thermodynamic equilibrium in the shelf region).  The shelf region erodes away throughout the implosion, until $r_h=r_g$, due to thermal transport from the hot spot to the shelf.  The shelf region is only present if the fuel is preheated from $r=0$ to $r<r_g$.}
\end{figure}

To elucidate some of the key physics issues relevant to MagLIF, a semi-analytic model of the concept was developed and verified.\cite{McBride_PoP_2015} This model is formulated as a system of ordinary differential equations (ODEs) that are straightforward to solve with standard software tools such as MATLAB\textsuperscript{\textregistered}, IDL\textsuperscript{\textregistered}, and Mathematica\textsuperscript{\textregistered}.   An overview of the model is provided in Fig.~\ref{fig:schematic}.  This model accounts for: (1) preheat of the fuel (optionally via laser absorption); (2) pulsed-power-driven liner implosion; (3) liner compressibility with an analytic equation of state, artificial viscosity, internal magnetic pressure, and ohmic heating; (4) adiabatic compression and heating of the fuel; (5) radiative losses and fuel opacity; (6) magnetic flux compression with Nernst thermoelectric losses; (7) magnetized electron and ion thermal conduction losses; (8) end losses; (9) enhanced losses due to prescribed dopant concentrations and contaminant mix; (10) deuterium-deuterium and deuterium-tritium primary fusion reactions for arbitrary deuterium to tritium fuel ratios; and (11) magnetized $\alpha$-particle heating.  This model has been implemented in a code called SAMM (Semi-Analytic MagLIF Model).  Simulations using SAMM typically take about 30 seconds to run on a laptop using the ode23 solver in MATLAB\textsuperscript{\textregistered}.  Using a parallel computing cluster at Sandia, parameter scans of about 2000 simulations can be completed in as little as 10 minutes.

In this paper, we present SAMM simulation results that: (a)~illustrate the parameter space, energetics, and overall system efficiencies of MagLIF; (b)~demonstrate the dependence of radiative loss rates on the radial fraction of the fuel that is preheated; (c)~explore some of the recent experimental results of the MagLIF program at Sandia;\cite{Gomez_PRL_2014} (d) highlight the experimental challenges presently facing the MagLIF program; and (e) demonstrate how increases to the preheat energy, fuel density, axial magnetic field, and drive current could affect future MagLIF performance.  Here, as in the original MagLIF paper (Ref.~\onlinecite{Slutz_PoP_2010}), we consider only standard gas-burning MagLIF.  That is, we do not consider the burning of a cryogenic DT ice layer, as in the ``high-gain'' MagLIF concept of Ref.~\onlinecite{Slutz_PRL_2012}; high-gain MagLIF is presently beyond the scope of SAMM.

\section{\label{sec2}Exploring the parameter space of MagLIF on the Z accelerator}

In Ref.~\onlinecite{Slutz_PoP_2010}, a preliminary point design for MagLIF on the Z accelerator is presented.  We will refer to this design as the ``2010 point design", which calls for the following: (1) a beryllium (Be) liner with an initial outer radius $r_{l0}=3.48$~mm, an axial length $h=5$~mm, and an initial aspect ratio $A_{r0}\equiv r_{l0}/(r_{l0}-r_{g0})=6$, where $r_{g0}$ is the initial radius of the fuel-liner interface; (2) a peak drive current of 27 MA; (3) an initial 50/50\% DT fuel density of 3 mg/cc; (4) an average preheated fuel temperature of 250 eV (which corresponds to about 8 kJ of preheat energy for the specified volume and density of the fuel); (5) uniformly preheated fuel with a preheat radius $r_{ph0}=r_{g0}$; and (6) an initial axial magnetic field $B_{z0}=30$ T embedded in the fuel.  In Ref.~\onlinecite{Slutz_PoP_2010}, the 1D simulation yield for this design is about 500 kJ and the convergence ratio $C_{rb}\equiv r_{g0}/r_{g,min}=25$ (with SAMM, we get about 970~kJ and $C_{rb}=25$).  

In Refs.~\onlinecite{Gomez_PRL_2014,Schmit_PRL_2014}, the results of the first fully integrated MagLIF experiments are presented.  The target design\cite{Sefkow_PoP_2014} was scaled down from the 2010 point design to better match the state of the Z facility at the time of these first experiments.  We will refer to these experiments as the ``2014 experiments", which consisted of the following: (1) an $A_{r0}=6$ Be liner with $r_{l0}=2.79$~mm, $r_{g0}=2.325$~mm, and $h=7.5$~mm; (2) a peak drive current of about 18~MA; (3) an initial DD fuel density of 0.7--1.5 mg/cc; (4) a laser energy of about 2 kJ delivered to the target (which may have deposited as little as 100--300 J in the fuel,\cite{Sefkow_PoP_2014} thus resulting in an average preheated fuel temperature of only about 8 eV); (5) $r_{ph0}=r_b\approx 500$~$\mu$m, where $r_b$ is the approximate radius of the laser beam in the fuel;\cite{rbcorrection} (6) $B_{z0}=10$ T; (7) an aluminum (Al) laser entrance channel (LEC) with a radius $r_{LEC}=1.5$~mm and an axial length $\Delta z_{LEC}=2.1$~mm; the LEC resides above the imploding liner region (see Fig.~2 in Ref.~\onlinecite{Gomez_PRL_2014}); (8) a laser entrance window (LEW) made from a 3.4-$\mu$m-thick polyimide foil; the LEW resides at the top of the LEC; and (9) a beam dump with a radius $r_{dump}=1$~mm; the beam dump resides below the imploding liner region (see Fig.~2 in both Refs.~\onlinecite{Gomez_PRL_2014} and \onlinecite{Gomez_PoP_2015}).  These experiments resulted in yields of up to $2\times 10^{12}$ DD neutrons.\cite{Gomez_PRL_2014} Also, spectroscopic analysis revealed the presence of contaminants mixed into the fuel.\cite{Hansen_PoP_2015} This contaminant ``mix'' was on the level of either 5--10\% Be (by atom) from the liner or 0.03--0.06\% Al (by atom) from the LEC.  The analysis was not able to discriminate between these two possible sources or determine when the mix occurred (i.e., if it occurred during laser preheating or later on in the implosion); however, more recent experiments support the hypothesis that the mix was largely laser induced.\cite{Gomez_PoP_2015}

The operating regime surrounding the 2010 point design is quite different from that surrounding the 2014 experiments.  To illustrate some of these differences, we present SAMM simulation results for both cases as well as some intermediate cases that we may be able to study experimentally in the near future.  We note that these scoping studies using SAMM are primarily for illustrative purposes; they should be considered merely as complementary to more sophisticated efforts presently underway using full 1D, 2D, and 3D radiation magnetohydrodynamics codes.

\subsection{\label{sec:2010pd}Further exploring the 2010 point design with SAMM: overall machine efficiency, preheating only a central portion of the fuel, and lithium liners}

\begin{figure*}
\includegraphics[width=0.75\textwidth]{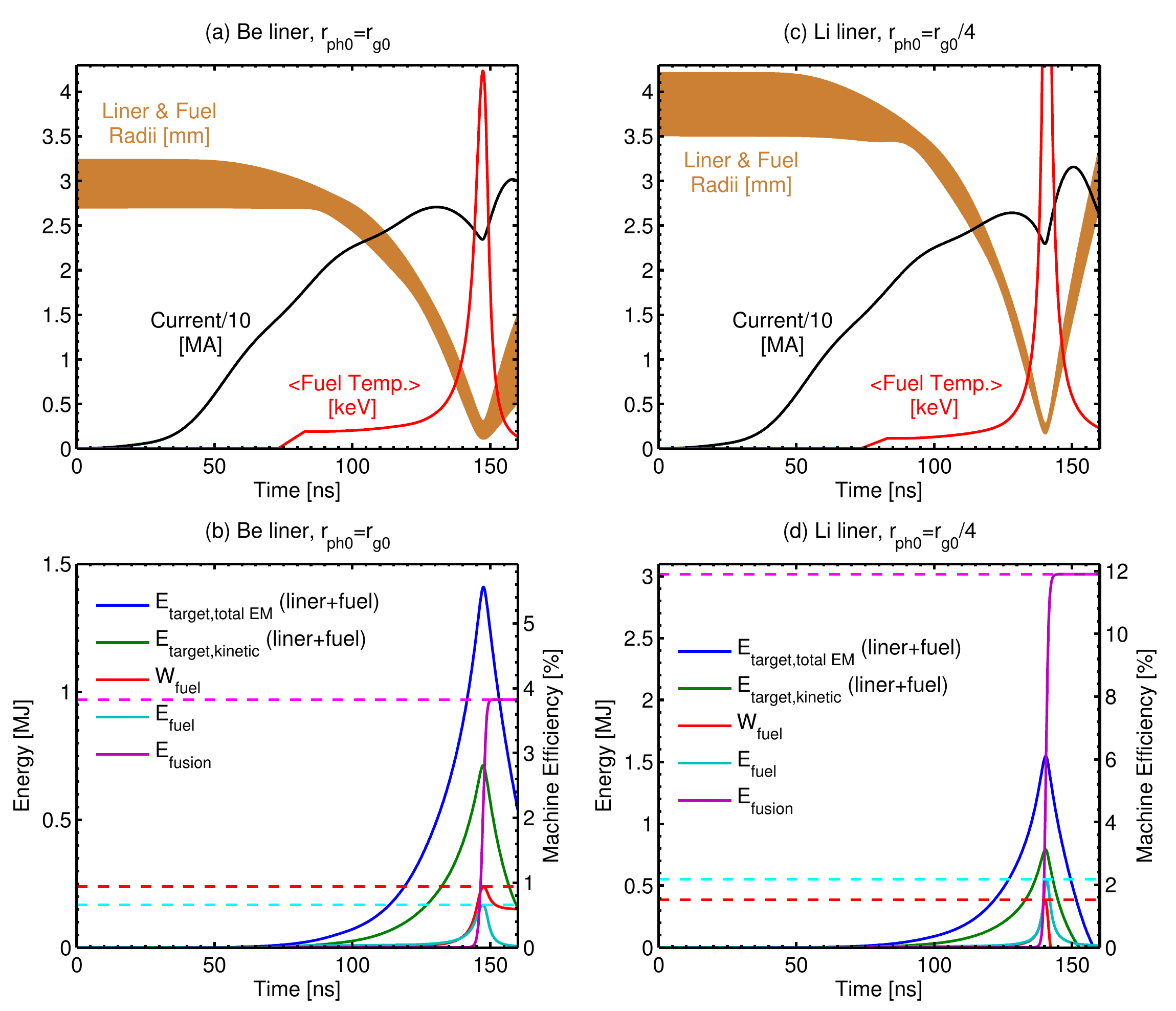}
\caption{\label{efficiency_montage} (a)~Implosion trajectory, drive current, and average fuel temperature for the 2010 point design. (b)~Energy balance and overall machine efficiency for the 2010 point design. (c)~Implosion trajectory, drive current, and average fuel temperature for the 2010 point design modified by using a lithium (Li) liner and by setting the preheat radius to one quarter of the initial fuel radius. (d)~Energy balance and overall machine efficiency for the simulation discussed in (c).}
\end{figure*}

In Fig.~\ref{efficiency_montage}(a), we show the 1D implosion dynamics of the 2010 point design.  Plotted are the liner and fuel radii, the electrical current that drives the implosion, and the spatially-averaged fuel temperature.  These dynamics were described in detail in Ref.~\onlinecite{Slutz_PoP_2010}.  Here, in Fig.~\ref{efficiency_montage}(b), however, we add to the description by presenting the energy balance and overall machine efficiency of the 2010 point design.  Since the point design calls for a peak drive current of 27~MA, the required Marx charge voltage for the Z accelerator is 95~kV, and thus the stored electrical energy is about 25~MJ.  The total electromagnetic energy delivered to the target (liner plus fuel) is just under 1.4 MJ (or about 5.5\%).  About half of the delivered electromagnetic energy is converted into kinetic implosion energy, and most of this is associated with the liner.  The work done on the fuel is about 250 kJ (or about 1\%).  Due to the accumulation of radiative and thermal conduction losses, the energy remaining in the fuel at stagnation is about 160 kJ (or about 0.65\%).  The fusion yield is calculated to be about 970 kJ. We find that about 0.8\% of the fuel is burned and that the Lawson criterion $p\tau=16$~Gbar$\cdot$ns, where this criterion is the product of the stagnation pressure $p$ and the confinement time $\tau$, and where we used the peak fuel pressure for $p$ and the burn pulse's full-width at half maximum (FWHM) for $\tau$.\cite{Betti_PoP_2010} Following Ref.~\onlinecite{Betti_PoP_2010} further, it is also useful to calculate $p\tau$ with alpha particle energy deposition turned off in a simulation.  This is because the implosion hydrodynamics at stagnation can be substantially affected by the increase in temperature and pressure that can occur in cases with intense heating by alpha particle energy deposition. For the case of the 2010 point design, we get $p\tau_{({\text{no-}}\alpha)}=12$~Gbar$\cdot$ns.

Some interesting modifications that could be made to the 2010 point design include the use of a lithium (Li) liner (rather than a Be liner) and setting the preheat radius to one quarter of the initial fuel radius (rather than heating all of the fuel uniformly).  Applying either one of these modifications on their own results in a doubling of the 2010 point design yield, while the application of both results in a yield tripling.  The energy balance and overall machine efficiency for applying these modifications is presented in Fig.~\ref{efficiency_montage}(d).  Note that the energy remaining in the fuel at stagnation exceeds the work done on the fuel, thus indicating that $\alpha$-heating dominates the radiative and thermal conduction losses in this case. Also, approximately 1.5\% of the fuel is burned, $p\tau=14$~Gbar$\cdot$ns, and $p\tau_{({\text{no-}}\alpha)}=11$~Gbar$\cdot$ns.  We note that, in general, MagLIF systems with 50/50\% DT fuel tend to self heat when $p\tau\gtrsim 10$~Gbar$\cdot$ns. This threshold is somewhat lower than in traditional ICF due to the enhanced alpha particle energy deposition that results from having a strong, flux-compressed axial magnetic field embedded in the fuel of a MagLIF implosion.

One of the reasons that Li liners enhance MagLIF performance is because of lithium's low mass density, which allows the liner to be scaled to larger radial dimensions while keeping the implosion time and the liner's initial aspect ratio fixed.  This results in higher implosion velocities and thus better coupling of machine energy to implosion kinetic energy.  Another reason that Li liners perform well in these calculations is because of lithium's compressibility.  For example, compare the Li implosion trajectory presented here in Fig.~\ref{efficiency_montage}(c) with the Be implosion trajectory presented in Fig.~\ref{efficiency_montage}(a).  The much more compressed lithium liner results in a comparable liner density at stagnation [$\rho_{Li}(t_{stag})\approx 0.8\rho_{Be}(t_{stag})$], despite lithium's much lower initial density [$\rho_{Li}(t_{0})= 0.29\rho_{Be}(t_{0})$].  Thus, by using Li, the implosion velocity can be substantially increased without excessively lowering the density of the inertially confining liner.  It should also be noted, however, that the higher in-flight aspect ratio (IFAR) of the lithium implosion could be detrimental to MagLIF with regards to the magneto-Rayleigh-Taylor instability and how quickly this instability can feed through to the liner's inner/fuel-confining surface.

The reasons why preheating only the central portion of the fuel enhances MagLIF performance were discussed in detail in Ref.~\onlinecite{McBride_PoP_2015}.  In short, rapidly preheating only the central portion of the fuel causes the preheated fuel to expand radially outward, pushing colder non-preheated fuel up against the liner's inner surface.  This results in the formation of two characteristic regions within the fuel: (I) a low-density hot spot region; and (II) a cold dense shelf region; ({\it cf.} Fig.~\ref{fig:schematic}).  This fuel structure has several advantages: (1) for a given amount of preheat energy, having a lower central density means that the central fuel temperature will be higher, and thus more reactive; (2) lowering the hot spot density reduces bremsstrahlung radiation losses from the hot fuel; and (3) the cold dense shelf region provides a buffer between the hot fuel and the cold liner wall that reduces radiative losses, thermal conduction losses, and magnetic flux losses from the fuel to the liner.  The presence of a cold dense shelf region reduces radiative losses because the fuel and the radiation field are in thermal equilibrium in the shelf region.  The presence of a cold dense shelf region reduces thermal conduction and magnetic flux losses because the temperature profile is nearly flat in the shelf region, and thus the temperature gradient to drive these transport mechanisms is small.

The presence of two characteristic fuel regions is not necessarily permanent throughout the implosion.  The cold dense buffer region begins to erode away immediately after its formation due to thermal conduction from the hot spot region to the shelf region ({\it cf}. Sec.~II~N in Ref.~\onlinecite{McBride_PoP_2015}).  In some cases (depending on the ratio of the preheat radius to the total fuel radius), the shelf region can completely erode away early on in the implosion.  Thus, in these cases, preheating the central portion of the fuel only delays the onset of more significant radiative and thermal conduction losses, and thus the performance gains are marginal.  By contrast, in Sec.~\ref{sec:2014exps}, we will see that for the 2014 MagLIF experiments, the shelf region is calculated to be present throughout the entire implosion, including stagnation and peak fusion burn.  The same is true for the modified 2010 point design presented here in Figs.~\ref{efficiency_montage}(c) and \ref{efficiency_montage}(d).

\subsection{\label{sec:2014exps}SAMM simulations of the 2014 MagLIF experiments}

In Ref.~\onlinecite{Gomez_PRL_2014}, four fully integrated MagLIF experiments (i.e., experiments that included both laser preheating and an applied axial magnetic field) are presented.  All of these experiments used DD fuel.  The DD neutron yields were $0.5\times 10^{12}$, $1\times 10^{12}$, and $2\times 10^{12}$ for the three experiments that used an initial fuel density of about 0.7 mg/cm$^3$.  For the one experiment that used an initial fuel density of 1.5 mg/cm$^3$, the the DD neutron yield was about $5\times 10^{9}$.  As discussed in both Refs.~\onlinecite{Sefkow_PoP_2014} and \onlinecite{Gomez_PRL_2014}, some of the biggest unknowns in these experiments are related to how much of the laser energy got past the LEW, how much of this transmitted energy coupled to the fuel, and where the remainder of the energy was deposited (i.e., whether or not some laser energy was backscattered and/or deposited in liner or electrode material).

The results of detailed 2D radiative magnetohydrodynamics simulations that varied the initial laser energy were presented in Fig.~12 of Ref.~\onlinecite{Sefkow_PoP_2014}.  They showed that, for the experiments with an initial fuel density of 0.7 mg/cm$^3$, the experimental data could be well modeled assuming a preheat energy of about 100--200~J.

Preheat levels on the order of 100 J are not unrealistic.  Even though the Z beamlet laser delivered about 2 kJ to the target in these first fully integrated experiments, subsequent laser-only experiments have found that $\sim$500~J or less gets transmitted through foils with materials and thicknesses similar to those used for the LEWs in the first fully integrated MagLIF experiments.\cite{Geissel_DPP_2014,laser_improvement_note}

\begin{figure}
\includegraphics[width=0.9\columnwidth]{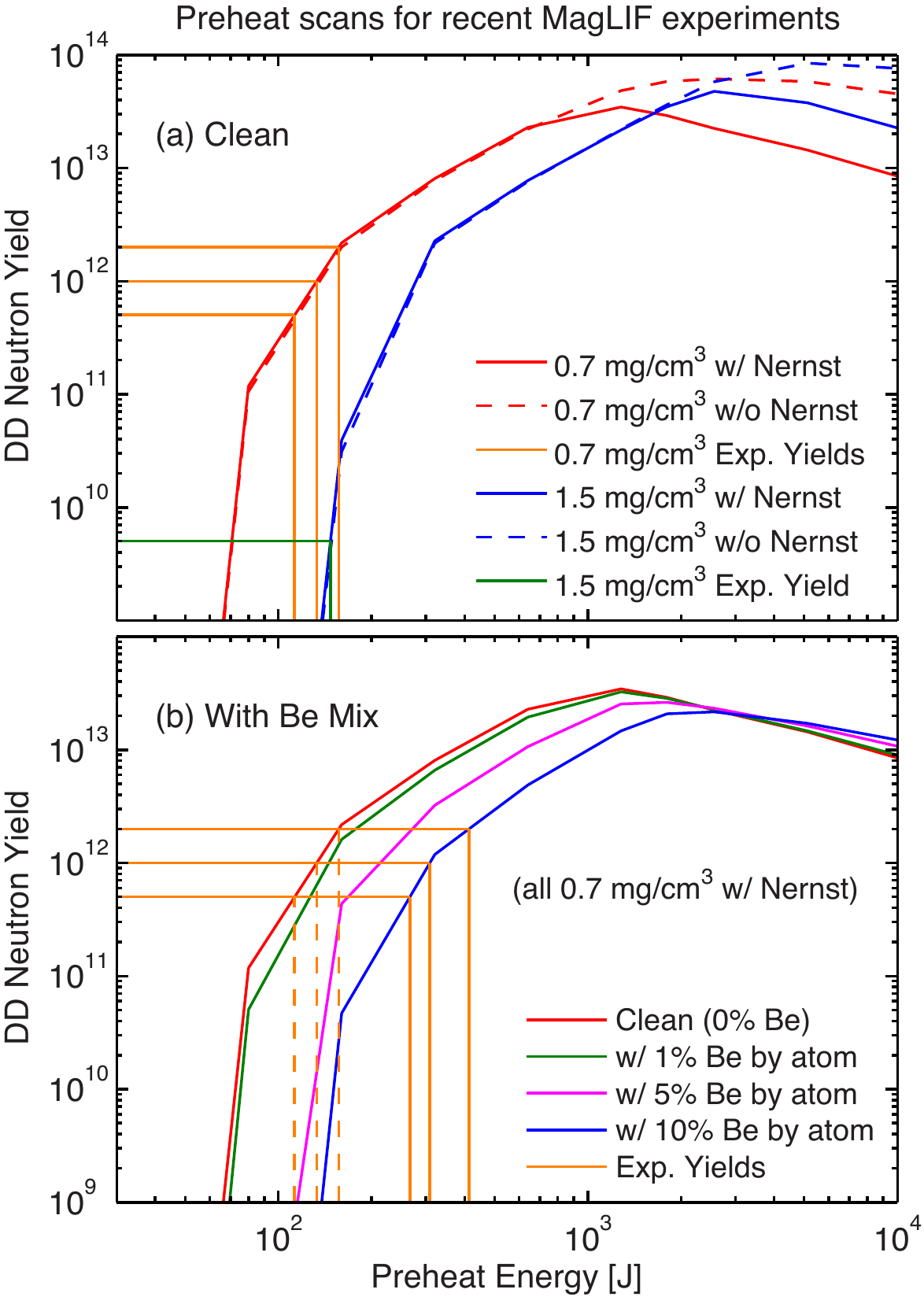}
\caption{\label{present_experiments_scan}Results from SAMM simulations of the fully-integrated MagLIF experiments presented in Ref.~\onlinecite{Gomez_PRL_2014}, where the preheat energy has been scanned over to find what preheat energy absorbed by the fuel results in DD neutron yields that are consistent with those measured in the experiments of Ref.~\onlinecite{Gomez_PRL_2014}.  (a) Clean scans, both with and without the Nernst effect included, that are comparable to those presented in Fig.~12 of Ref.~\onlinecite{Sefkow_PoP_2014}. (b) Scans with various levels of Be mix for the 0.7 mg/cm$^3$ case with Nernst.}
\end{figure}

Here, in Fig.~\ref{present_experiments_scan}, we present the results from preheat energy scans using SAMM, where we used the 2014 experimental parameters listed at the beginning of Sec.~\ref{sec2}, including a preheat radius $r_{ph0}=r_b\approx 500$ $\mu$m.  In Fig.~\ref{present_experiments_scan}(a), the red (blue) curves were generated using an initial fuel density of 0.7~mg/cm$^3$ (1.5 mg/cm$^3$), while the solid (dashed) lines represent scans with the Nernst effect\cite{Braginskii_1965,Slutz_PoP_2010,Velikovich_PoP_2015,McBride_PoP_2015} turned on (off).  Also, the orange (green) horizontal lines represent the measured DD neutron yields for the experiments with an initial fuel density of 0.7 mg/cm$^3$ (1.5 mg/cm$^3$), while the corresponding vertical lines indicate the prescribed preheat energy required to obtain these yields in the SAMM calculations.  These scans using SAMM give very similar results to those presented in Ref.~\onlinecite{Sefkow_PoP_2014}; namely, that the measured experimental yields could be explained by only 110--160~J of laser energy coupling to the imploding region of the fuel.  These results are also consistent with roughly 160--330~J making it past the LEW, as 40--190~J is absorbed by the fuel above the imploding region, in the 2.1-mm-long LEC.\cite{lecapproxc,laserpropdopcorrection}  Moreover, these results suggest that the Nernst effect is not a significant factor in these experiments. 

The amount of preheat energy inferred from Fig.~\ref{present_experiments_scan}(a) is not unique, and it is sensitive to the level of mix assumed. This is illustrated by the simulation results presented in Fig.~\ref{present_experiments_scan}(b), where the fuel has been uniformly premixed with various levels of Be (note that 5--10\% Be is the amount inferred for the 2014 experiments\cite{Gomez_PRL_2014,Hansen_PoP_2015}).  From the 10\% curve, we see that as much as $\sim$400~J could have coupled to the fuel in the 2014 experiments.  Additionally, given that $\sim$100~J is absorbed in the LEC above the imploding region, the total laser energy that got past the LEW could have been $\sim$500~J, which would agree well with laser-only foil transmission experiments.\cite{Geissel_DPP_2014}

\begin{figure*}
\includegraphics[width=0.9\textwidth]{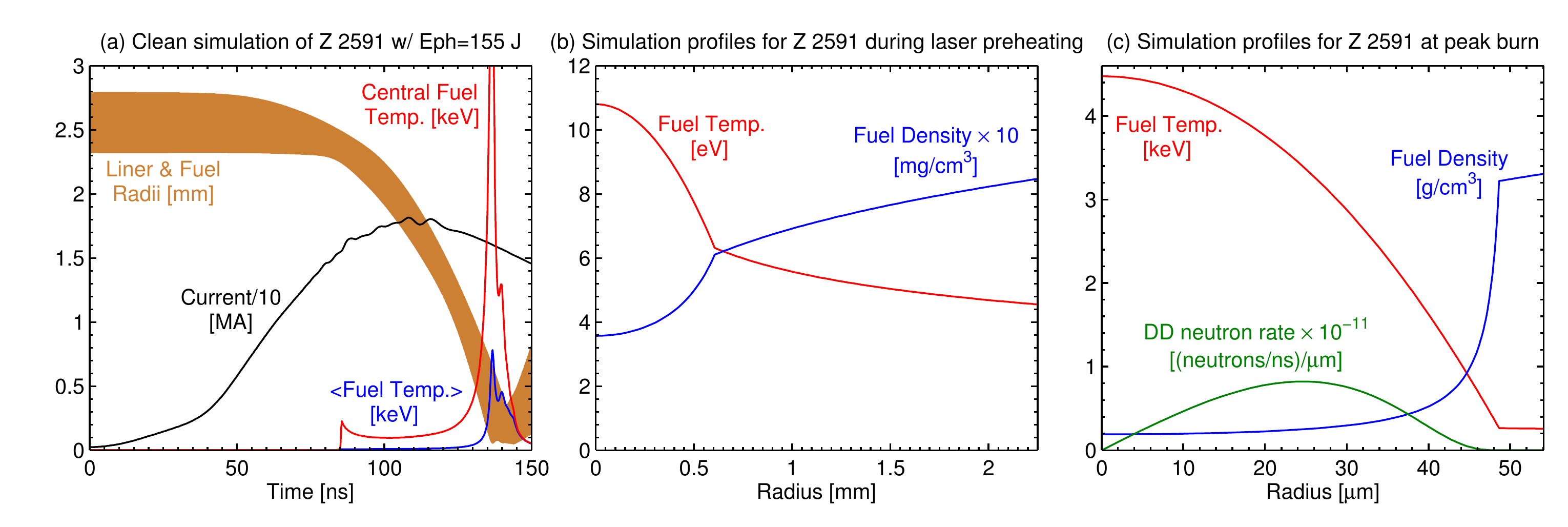}
\caption{\label{present_experiments_montage}Results from a clean SAMM simulation of Z shot 2591 (i.e., the best performing fully-integrated MagLIF experiment presented in Ref.~\onlinecite{Gomez_PRL_2014}). (a)~Liner and fuel implosion trajectories (orange), experimentally measured load current used to drive the simulation (black), average fuel temperature (blue), and central (on-axis) fuel temperature (red).  With a prescribed preheat energy of 155 J, the simulated yield matches the experimentally measured yield of $2\times 10^{12}$ DD neutrons.\cite{Gomez_PRL_2014} (b)~Radial profiles for the fuel temperature and fuel density 1.5 ns into the 2-ns laser preheating pulse. (c)~Radial profiles for the fuel temperature, fuel density, and DD neutron production rate at the time of peak fusion burn.}
\end{figure*}

In Fig.~\ref{present_experiments_montage}, we present results from a clean SAMM simulation of the best performing MagLIF experiment of Ref.~\onlinecite{Gomez_PRL_2014} (i.e., Z shot 2591, which yielded $2\times 10^{12}$ DD neutrons). This simulation was driven directly by the measured experimental load current, plotted in black in Fig.~\ref{present_experiments_montage}(a).  Although Z shot 2591 is the best performing MagLIF experiment of Ref.~\onlinecite{Gomez_PRL_2014}, it is still considered a preheat-energy-starved case if we assume that the preheat energy was only about 155 J, which is what is required for the SAMM simulation to give the experimentally measured yield of $2\times 10^{12}$ DD neutrons. 

The results presented in Fig.~\ref{present_experiments_montage} illustrate the consequences of the low preheat energy and how such a low preheat energy case can still obtain a decent yield of $2\times 10^{12}$ DD neutrons.  For example, from Fig.~\ref{present_experiments_montage}(a), we see that the spatially-averaged fuel temperature (blue) peaks at only about 780 eV; however, what enables significant fusion reactions to occur is the fact that the hot spot temperature is considerably higher, peaking at over 4 keV on axis.  This is represented by the central (on-axis) fuel temperature plotted in red as a function of time in Fig.~\ref{present_experiments_montage}(a) and by the fuel temperature's radial profile plotted in red at the time of peak fusion burn in Fig.~\ref{present_experiments_montage}(c).

The reason the central temperature is so much higher than the average fuel temperature is because the preheat radius was only about 21\% of the initial fuel radius ($r_{ph0}=r_b\approx 500\,\mu{\rm m}\approx 0.21r_{g0}$), which establishes a hot spot region (initially with only 4\% of the fuel mass) and a cold high density shelf region (initially with 96\% of the fuel mass).  This initial state is represented by the fuel profiles plotted in Fig.~\ref{present_experiments_montage}(b).  Because the shelf region contained so much of the fuel mass initially (at the time of preheating), and because the preheat energy was so low (155~J), erosion of this massive shelf region was slow enough for the shelf to persist throughout the implosion and beyond peak burn.  Note that the presence of the shelf region is still clearly visible in Fig.~\ref{present_experiments_montage}(c), which is at the time of peak burn.

This simulation also accounted for the loss of fuel mass (and the energy carried by the fuel mass) from the imploding region through the top and bottom apertures formed by the LEC and the beam dump, respectively (see Sec.~II~L in Ref.~\onlinecite{McBride_PoP_2015} for details on this end-loss model).  At the time of peak fusion burn, 84\% of the initial fuel mass remains in the imploding region, of which, 57\% is in the shelf region and 43\% is in the hot spot region.

From the DD neutron production rate, plotted in green in Fig.~\ref{present_experiments_montage}(c), we see that essentially all of the DD neutrons come from the hot spot region (and thus $\lesssim$43\% of the fuel mass contributes to their production).  Also note that the peak in the DD neutron rate corresponds to a fuel temperature of about 3.3 keV, which is in agreement with the experiments of Ref.~\onlinecite{Gomez_PRL_2014}, where the measured ion temperatures were 2.5$\pm$0.8~keV (from DD neutron time-of-flight data) and the measured electron temperatures were $3.1^{+0.7}_{-0.5}$~keV (from x-ray spectroscopy data).\cite{Gomez_PRL_2014}

In the simulation of Z shot 2591, $p\tau=0.6$~Gbar$\cdot$ns, and of course $p\tau=p\tau_{({\text{no-}}\alpha)}$ since DD fuel was used. Also, the fuel pressure at stagnation is calculated to be about 0.8 Gbar, which is in good agreement with the 0.9-Gbar stagnation pressure inferred experimentally for Z shot 2591.\cite{Hansen_PoP_2015}  Additionally, this pressure is comparable to the 3--6 Gbar peak pressure of the 2010 point design.  However, for the Z shot 2591 simulation to obtain this pressure, the convergence ratio for the fuel-liner interface was about 43 at peak burn (as opposed to the 2010 point design's convergence ratio of only 25).  Note that this somewhat lower stagnation pressure and higher convergence ratio affect the liner implosion trajectory near stagnation, where the liner's inner surface can undergo multiple bounces.  For example, compare the weak stagnation bounce(s) for the energy-starved case presented here in Fig.~\ref{present_experiments_montage}(a) with the much stronger bounces presented above in Figs.~\ref{efficiency_montage}(a) and \ref{efficiency_montage}(c).  For energy-starved cases, we often find that peak fusion burn occurs well before maximum compression (e.g., in the Z shot 2591 simulation, the convergence ratio at maximum compression was about 47, and it occurred about 7 ns after the time of peak burn).  In Sec.~\ref{sec:Znext}, however, we will present examples of igniting cases where just the opposite happens; i.e., where peak fusion burn occurs after a strong bounce, during radial expansion, due to substantial $\alpha$-heating rates.  For reference, the 2010 point design has a peak fusion burn that occurs simultaneously with peak compression.

From Fig.~\ref{present_experiments_montage}(c), we see that the minimum radius for the fuel-liner interface is about 54 $\mu$m (which corresponds to the convergence ratio of 43 stated above), while the radius for the hot spot region is about 49 $\mu$m (due to the substantial thickness of the shelf region), and the radius for the peak neutron production rate is about 25 $\mu$m.  All of these radii are in reasonable agreement with the experimental data of Ref.~\onlinecite{Gomez_PRL_2014}.  For example, the bright helical-like column in the time-integrated image of Fig.~5 in Ref.~\onlinecite{Gomez_PRL_2014} was generated by $>$6-keV photons; therefore, it provides a reasonable surrogate for where the neutrons were likely produced.\cite{Hansen_PoP_2015}  This column has an axially-varying diameter in the range of 70--110 $\mu$m at full-width half-maximum (effectively 35--55 $\mu$m in radius).\cite{Gomez_PRL_2014}

The stagnation profiles presented in Fig.~\ref{present_experiments_montage}(c) are in good agreement with the stagnation profiles presented in Fig.~2 of Ref.~\onlinecite{Hansen_PoP_2015}, which were constrained by the experimental data of Ref.~\onlinecite{Gomez_PRL_2014}.  However, the profiles in Ref.~\onlinecite{Hansen_PoP_2015} included 5--10\% Be mix (by atom), whereas the simulation results presented here in Fig.~\ref{present_experiments_montage} assumed no mix. The agreement is reasonable, though, since the simulation's preheat energy is adjusted to obtain the experimentally measured DD neutron yield.  For example, simulations of Z shot 2591 with 5--10\% Be mix give results very similar to the clean results presented here, as long as the preheat energy is increased from 155~J to 250--400~J to compensate for the additive mix.

Further discussion of mix is beyond the scope of this paper.  However, we note that a better understanding of both laser-fuel coupling and mix is needed. In particular, we need to better understand how mix scales with preheat energy.  For example, if increasing the preheat energy increases mix too fast, then the increased preheat energy could actually degrade MagLIF performance. Indeed, laser-induced mix from the LEC is one hypothesis for the degraded yields measured on more recent MagLIF experiments, where thinner LEWs were used and thus greater laser coupling to the fuel has been assumed.\cite{Gomez_PoP_2015}

\begin{figure}
\includegraphics[width=0.9\columnwidth]{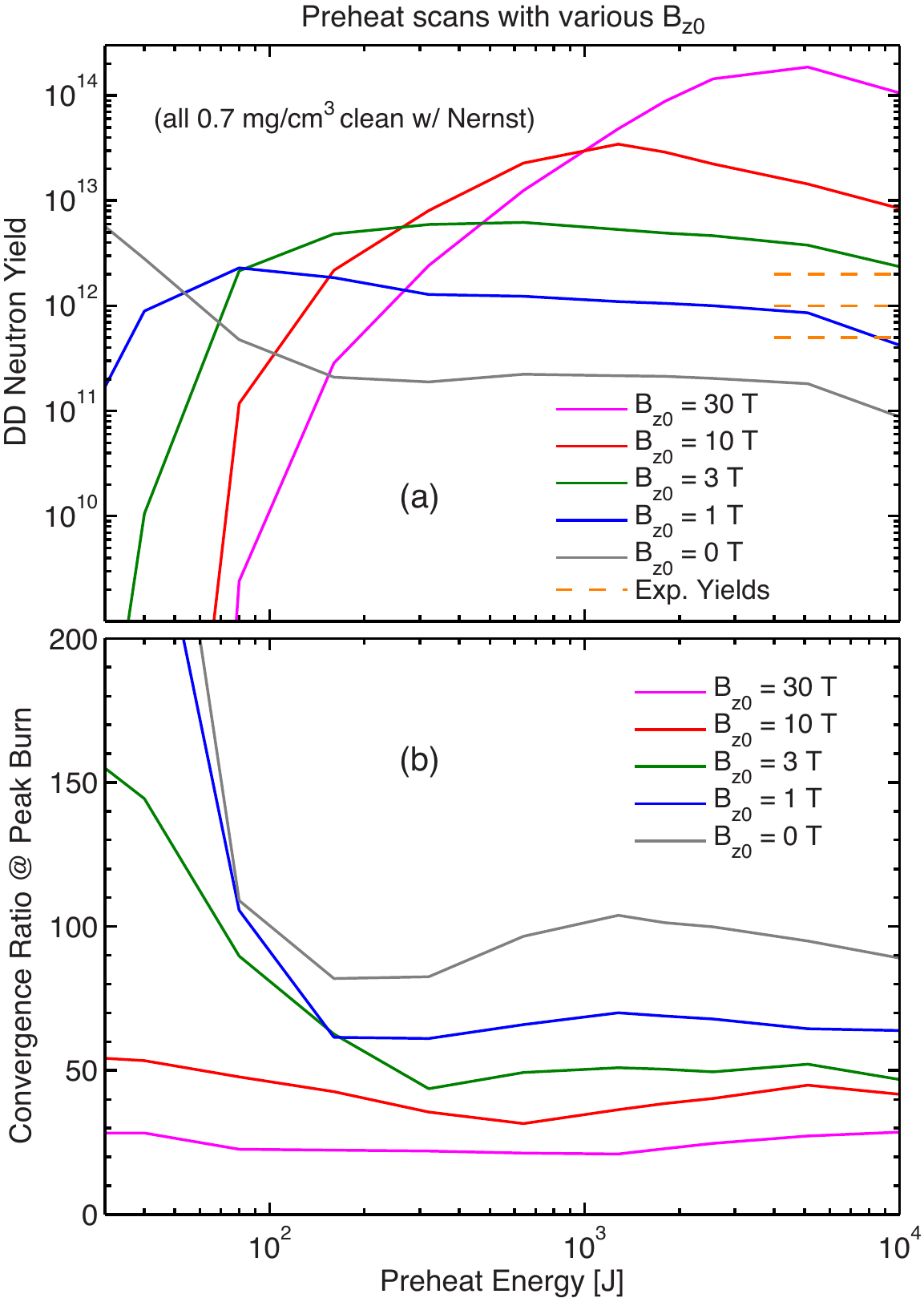}
\caption{\label{present_experiments_scan_varyBz0}Results from SAMM simulations of the fully-integrated MagLIF experiments presented in Ref.~\onlinecite{Gomez_PRL_2014}, where both the preheat energy and the initial axial magnetic field, $B_{z0}$, have been varied.  (a)~DD neutron yields. (b)~Convergence ratios at peak burn.}
\end{figure}

For the very near future, a series of experiments is planned where $B_{z0}$ will be varied. The experimental platform will be similar to the one used in the 2014 experiments.  To understand how the planned experiments may depend on $B_{z0}$, several SAMM simulations were run.  The results of these simulations are presented in Fig.~\ref{present_experiments_scan_varyBz0}. Curiously, the results in Fig.~\ref{present_experiments_scan_varyBz0}(a) suggest that, for very low preheat energies (below about 200~J), the yield could be increased by lowering $B_{z0}$.  However, these 1D results should be met with skepticism, since the required convergence ratios [Fig.~\ref{present_experiments_scan_varyBz0}(b)] are uncomfortably high in this regime and likely cannot be achieved uniformly in a real (3D) experiment.  In MagLIF, lower convergence ratios are sought after to reduce the risk of 3D implosion instabilities disrupting fusion burn.  The results presented in Fig.~\ref{present_experiments_scan_varyBz0} demonstrate that increasing $B_{z0}$ both reduces the risk (reduces the convergence ratios) and increases the yields, as long as significant preheat energies can be obtained (i.e., above 1~kJ).

\subsection{\label{sec:Z}SAMM parameter scans for MagLIF on the Z accelerator}

\begin{figure*}
\includegraphics[width=1\textwidth]{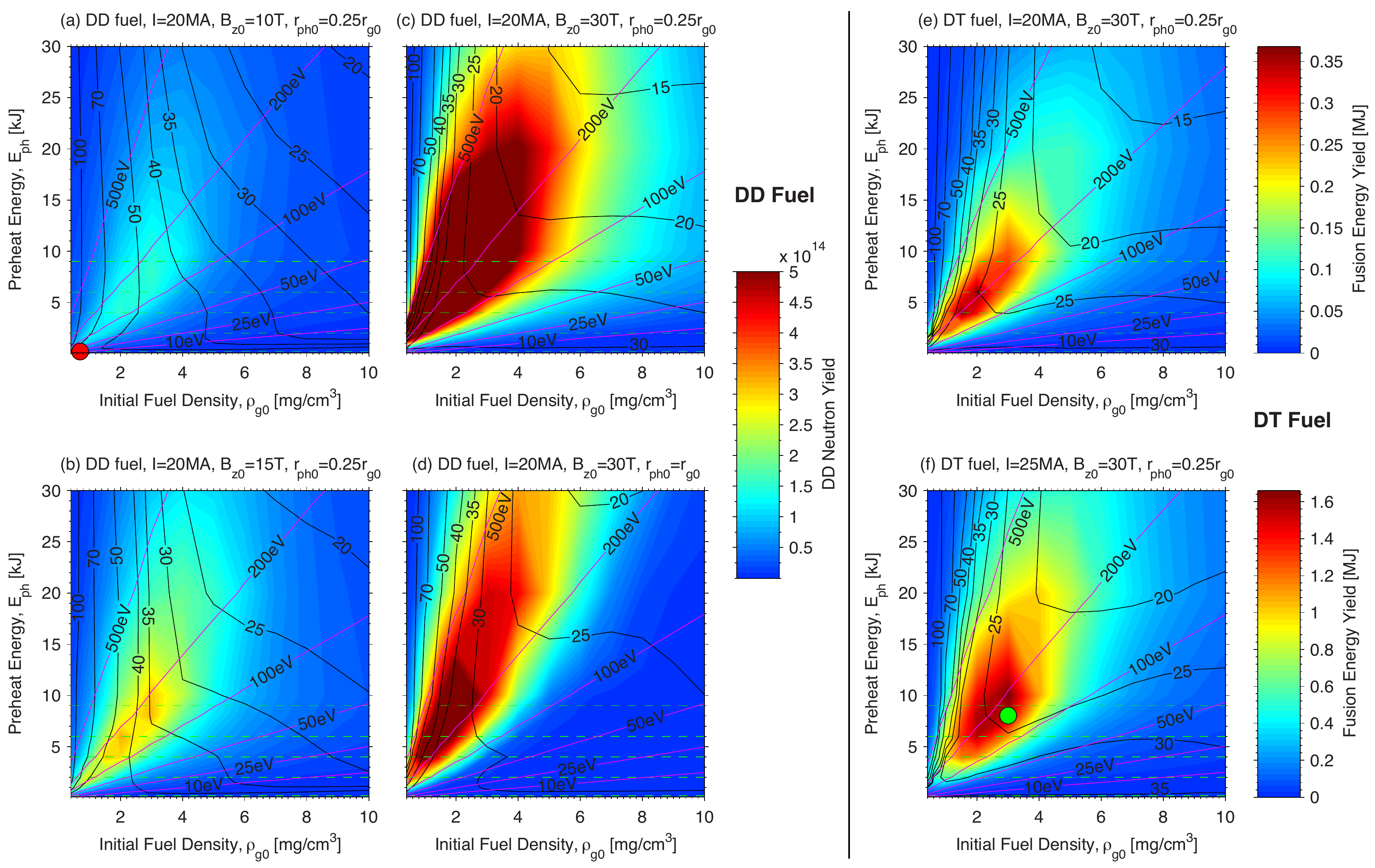}
\caption{\label{near_term_contour} SAMM parameter scan results for MagLIF on the Z accelerator. The overlaid black lines are contours of constant maximum convergence ratio.  The overlaid magenta lines are lines of constant preheat temperature.  The overlaid dashed green lines represent various laser preheat energies of interest.  The parameter space in (a) is similar to that surrounding the 2014 experiments (red circle).  The space in (f) is similar to that surrounding the 2010 point design (green circle). The spaces in (b)--(c) and (e)--(f) may be accessible in the near future with modest facility upgrades. The case in (d) is shown to illustrate the performance degradation [relative to (c)] that occurs when all of the fuel is preheated uniformly. For all of the simulations presented in (a)--(f), SAMM's end-loss model (described in Sec.~II~L of Ref.~\onlinecite{McBride_PoP_2015}) was turned on to simulate single-ended fuel loss through a hole with a radius equal to the preheat radius, i.e., $r_{hole}=r_{ph0}=0.25r_{g0}$.}
\end{figure*}

In an effort to see how various upgrades to the Z facility might affect future MagLIF performance, we have run several parameter scans using SAMM (see Fig.~\ref{near_term_contour}).  All of these simulations were clean and used 7.5-mm-long Be liners with an initial aspect ratio of 6.  For the simulations with a peak drive current of 20 MA (25 MA), the liner's initial outer radius was 2.8 mm (3.1 mm).  The results presented in Fig.~\ref{near_term_contour} show the effects of varying the initial fuel density (horizontal axis in the plots) and the deposited preheat energy (vertical axis in the plots).  The filled color contours represent either DD neutron yields [Figs.~\ref{near_term_contour}(a)--\ref{near_term_contour}(d)] or DT fusion energy yields [Figs.~\ref{near_term_contour}(e)--\ref{near_term_contour}(f)], depending on whether DD or DT fuel was used in the simulation.  The overlaid black lines are contours of constant maximum convergence ratio.  The overlaid diagonal magenta lines are lines of constant preheat temperature.  The overlaid dashed green lines at constant preheat energies of 150~J, 2~kJ, 4~kJ, 6~kJ, and 9~kJ, represent various laser preheat energies of interest.  That is, the 150-J line (essentially the bottom horizontal axis in these plots) represents what we think may have coupled to the fuel in the experiments of Ref.~\onlinecite{Gomez_PRL_2014}.  The 2-kJ line represents the beam energy that was delivered to the targets in the experiments of Ref.~\onlinecite{Gomez_PRL_2014}.  The 4-kJ line represents the beam energy that can be delivered to a target today (i.e., the laser has been upgraded from 2 kJ to 4 kJ since the time of the experiments in Ref.~\onlinecite{Gomez_PRL_2014}).  The 6-kJ line represents where we would like the laser to be in the next year or two, after some planned upgrades.  Finally, the 9-kJ line represents where the Z beamlet laser could potentially be with enhanced resources.

We note that the critical density for the frequency doubled Z beamlet laser (527 nm) is 13 mg/cm$^3$ for DD fuel and 17 mg/cm$^3$ for DT fuel. To avoid driving non-ideal laser-plasma interactions, it is best to keep the initial fuel density below 10\% of the critical density. Thus, to access the higher initial fuel densities shown in Fig.~\ref{near_term_contour}, a frequency tripled Z beamlet laser may be required. This system would raise the critical density to 30 mg/cm$^3$ for DD fuel and 38 mg/cm$^3$ for DT fuel. Alternatively, a KrF laser (248 nm) could be employed, which would raise the critical density to 61 mg/cm$^3$ for DD fuel and 76 mg/cm$^3$ for DT fuel. 

In all of the plots shown in Fig.~\ref{near_term_contour}, a characteristic optimal region is apparent.  This region expands outward from the lower left of each plot, toward the upper right of each plot.  Above and to the left of the optimal region (where there are high preheat energies and high preheat temperatures), MagLIF performance degrades primarily due to thermal conduction losses; it is in this region that the Nernst effect plays its largest role.  Below and to the right of the optimal region (where there are high initial fuel densities and low preheat temperatures), MagLIF performance degrades primarily due to the low preheat temperatures {\it and} due to increased bremsstrahlung losses; it is this region that is degraded the most when all of the fuel is preheated uniformly.  

Figure~\ref{near_term_contour}(a) best encompasses the experiments of Ref.~\onlinecite{Gomez_PRL_2014} (i.e., DD fuel, a peak drive current of 20~MA, an initial axial magnetic field of 10~T, and a preheat radius that is 25\% of the initial fuel radius).   Recently, the applied $B$ on Z (ABZ) subsystem at the Z facility has been upgraded to 15--20~T.  In the near future, we expect 30~T to be available.\cite{Rovang_RSI_2014}  Because of these upgrades, we have varied the initial magnetic field strength from 10~T, to 15~T, to 30~T in Figs.~\ref{near_term_contour}(a), \ref{near_term_contour}(b), and \ref{near_term_contour}(c), respectively.  From these plots, one can see that increasing the initial axial magnetic field not only enhances neutron yields, but also lowers the required convergence ratios substantially.

In Fig.~\ref{near_term_contour}(d), we plot results for preheating all of the fuel uniformly (rather than using a preheat radius that is 25\% of the initial fuel radius).  This figure should be compared with Fig.~\ref{near_term_contour}(c), since an initial axial magnetic field of 30 T was used for both cases.  One can see that preheating all of the fuel uniformly not only decreases the neutron yields, but also significantly reduces the size of the optimal parameter space; i.e., the use of higher initial fuel densities becomes undesirable due to enhanced bremsstrahlung losses.  This phenomenon (discussed briefly in Sec.~\ref{sec:2010pd} of this paper and in detail in Sec.~II~J of Ref.~\onlinecite{McBride_PoP_2015}) becomes even more pronounced as MagLIF is scaled to larger accelerators, as is demonstrated in Sec.~\ref{sec:Znext} below.

In Fig.~\ref{near_term_contour}(e), we plot results for changing the fuel from DD to DT.  This figure should be compared with Fig.~\ref{near_term_contour}(c), since the only change between the two cases is the fuel constituents.  Note that various performance metrics (e.g., the convergence ratio contours and the shapes of the optimal regions) are very similar for the two cases.  Also note that exceeding a fusion energy yield of about 100 kJ with DT fuel in Fig.~\ref{near_term_contour}(e) is roughly equivalent to exceeding a DD neutron yield of about 2--$4\times 10^{14}$ neutrons with DD fuel in Fig.~\ref{near_term_contour}(c). In these simulations, 100~kJ is roughly the amount of energy delivered to the fuel by the combined efforts of the preheating and the implosion. These scans suggest that exceeding 100 kJ of fusion energy yield from DT fuel and/or 2--$4\times 10^{14}$ DD neutrons from DD fuel might be possible with a number of different preheat energy and initial fuel density combinations, as long as the coupled preheat energy can be increased to $\gtrsim$ 2 kJ.

In Fig.~\ref{near_term_contour}(f), we present results for increasing the peak drive current to 25 MA and using DT fuel.  With these upgrades and the next planned upgrade to a 6 kJ laser, fusion energy yields $>100$~kJ may be robustly possible.  This parameter space is similar to that surrounding the 2010 point design.

\section{Exploring the parameter space of MagLIF on future pulsed-power facilities\label{sec:Znext}}

Two conceptual designs presently being studied that represent possible future pulsed-power accelerators are the Z300 and Z800 designs.\cite{Stygar_PRSTAB_2015}  The numbers ``300'' and ``800'' refer to the approximate peak electrical power (in TW) delivered to the experiment's vacuum section.  Both of these designs provide flexible pulse shaping capabilities and are based on the new and revolutionary pulsed power technology referred to as compact Linear Transformer Drivers (LTDs).\cite{Kim_PRSTAB_2009} LTD-based architectures are about 2--4 times more efficient than standard Marx-generator-based architectures.  For example, in a footprint roughly the size of today's 80-TW Z machine (35 m in diameter), the Z300 design consists of: (1) an electrical energy of 48 MJ stored in its capacitors prior to the experiment; (2) 320 TW delivered to the experiment's vacuum section; (3) 48 MA delivered to an experimental load; and (4) a rise time of 120--155 ns (0--48 MA).  In a slightly larger footprint (52 m in diameter), the Z800 design consists of: (1) an electrical energy of 130 MJ stored in its capacitors prior to the experiment; (2) 890 TW delivered to the experiment's vacuum section; (3) 60--65 MA delivered to an experimental load; and (4) a rise time of 110--120 ns (0--60 MA).



\begin{figure*}
\includegraphics[width=0.95\textwidth]{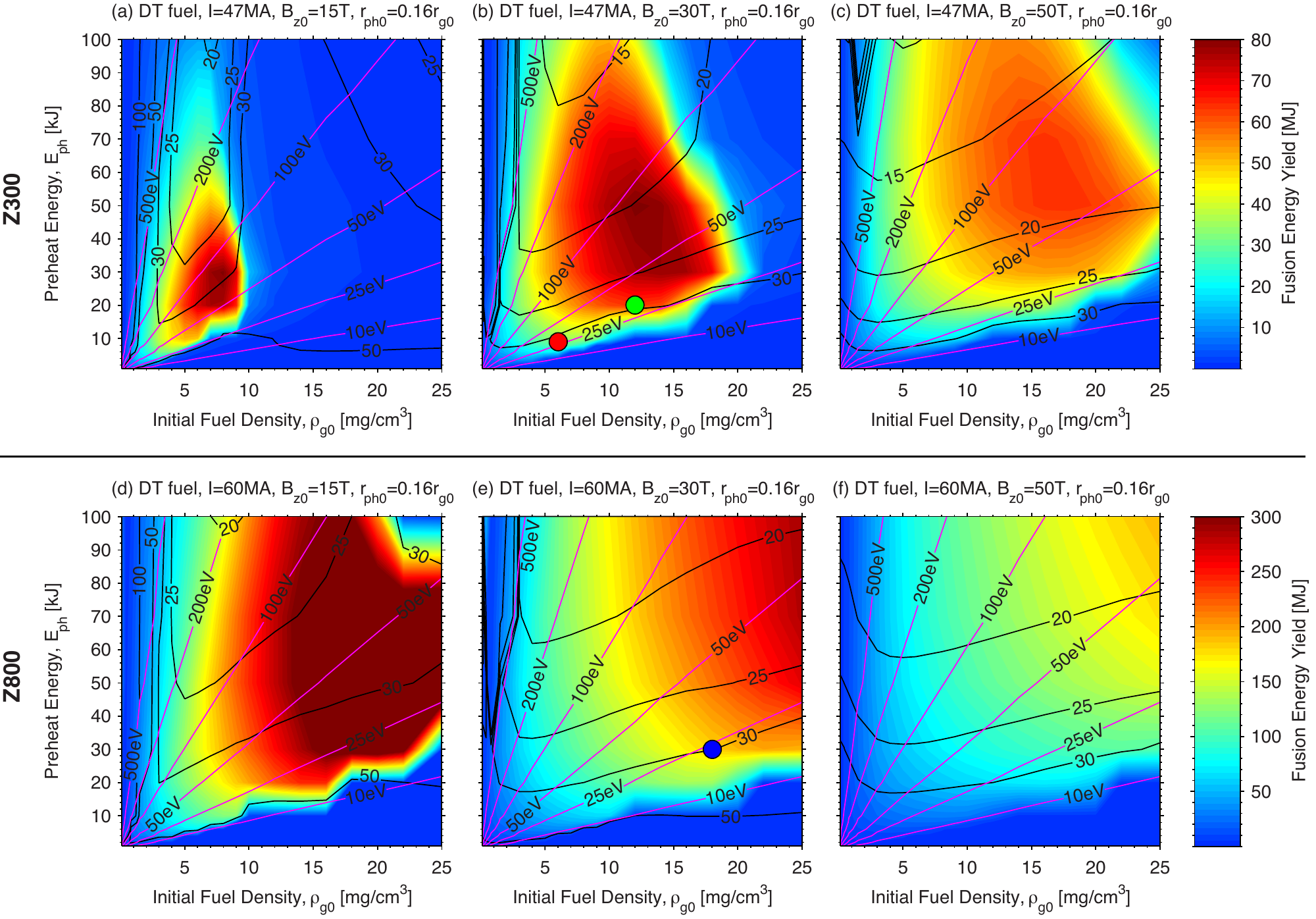}
\caption{\label{future_contours}SAMM parameter scan results for MagLIF on the conceptual Z300 (top row) and Z800 (bottom row) accelerators. The circles in (b) and (e) correspond to the cases presented in Fig.~\ref{Z300_Z800_rt_montage}. In all cases, the preheat radius was set to 16\% of the initial fuel radius and SAMM's end-loss model was turned off.}
\end{figure*}

Using SAMM, we have run several parameter scans using Z300 and Z800 drive circuit models.  All of these simulations were clean. The Z300 simulations used 9-mm-long Be liners with an initial outer radius of 4.6~mm and an initial aspect ratio of 6.  The Z800 simulations used 10-mm-long Be liners with an initial outer radius of 5 mm and an initial aspect ratio of 6.  The results of these scans are presented in Fig.~\ref{future_contours}. In the top row, we present the results for Z300, where we have set the initial axial field to 15 T, 30 T, and 50 T in Figs.~\ref{future_contours}(a), \ref{future_contours}(b), and \ref{future_contours}(c), respectively. Note that the optimal parameter space continues to broaden, and the convergence ratios continue to decrease, as we increase the initial axial magnetic field strength; however, the optimum yield appears in the 30-T case.  In the bottom row of Fig.~\ref{future_contours}, we present the results for Z800, where we have set the initial axial field to 15 T, 30 T, and 50 T in Figs.~\ref{future_contours}(d), \ref{future_contours}(e), and \ref{future_contours}(f), respectively. Note that the optimal parameter space broadens as we go from Z300 to Z800. For Z800, the optimum yield appears in the 15-T case. We again note that if the preheating is done with a laser, then a frequency tripled Nd:glass laser or a KrF laser will likely be required to access the higher initial fuel densities in these plots.

\begin{figure}
\includegraphics[width=0.9\columnwidth]{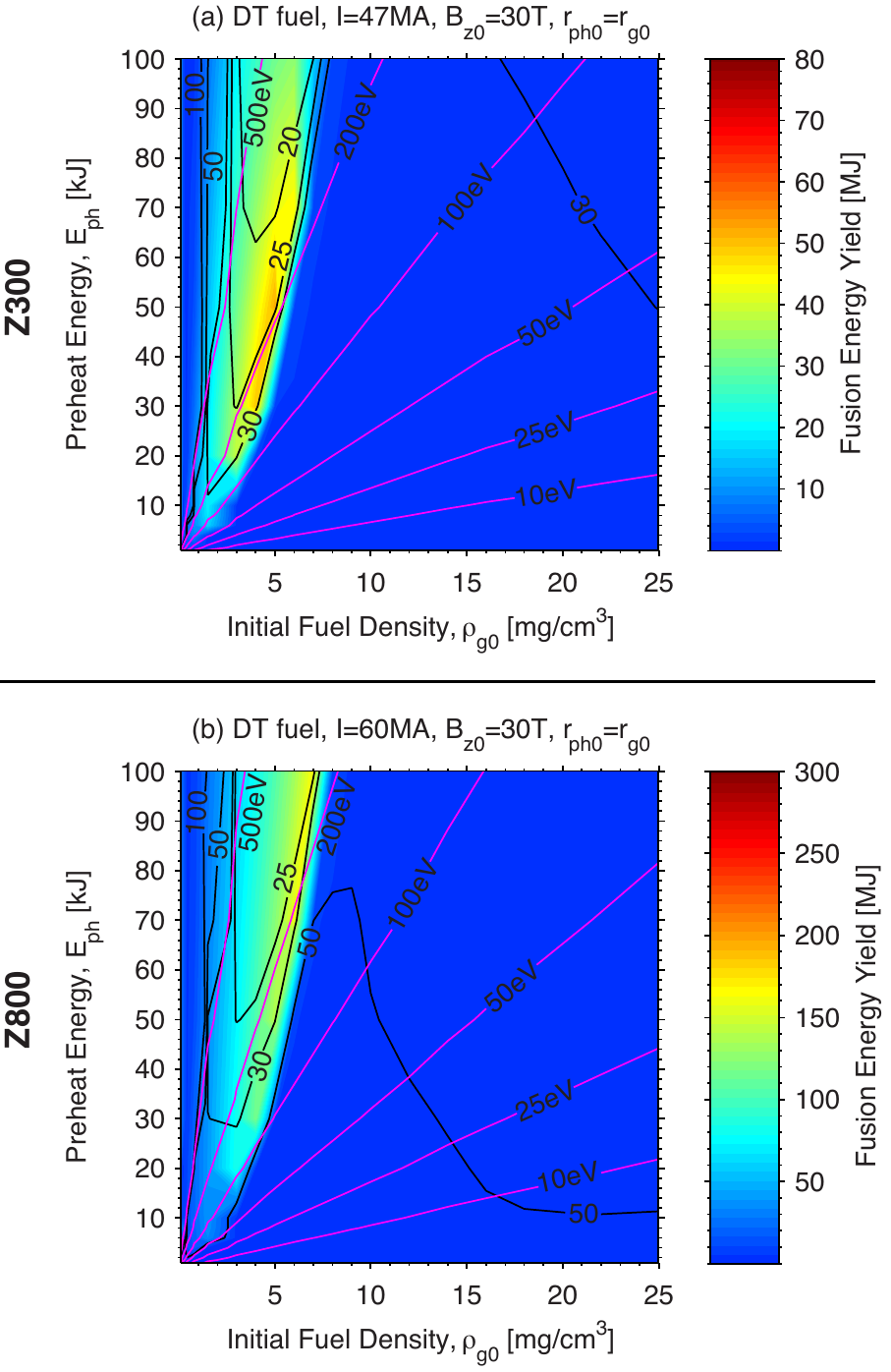}
\caption{\label{Z300_Z800_contour}SAMM parameter scan results for MagLIF on the conceptual Z300 (top) and Z800 (bottom) accelerators. For both cases, the preheat radius was set to 100\% of the initial fuel radius.  The results in (a) and (b) should be contrasted with those presented Figs.~\ref{future_contours}(b) and \ref{future_contours}(e), respectively.  These results are presented to illustrate the very pronounced performance degradation that occurs when all of the fuel is preheated uniformly at these higher drive currents.}
\end{figure}

For all of the results presented in Fig.~\ref{future_contours}, the fuel preheat radius was set to 16\% of the initial fuel radius.  As mentioned in Sec.~\ref{sec:Z}, the negative effects of preheating all of the fuel uniformly is more pronounced at higher drive currents.  To illustrate this point, we reran the scans of Figs.~\ref{future_contours}(b) and \ref{future_contours}(e) with the preheat radius set to 100\% of the initial fuel radius.  The results of these two scans are presented in Figs.~\ref{Z300_Z800_contour}(a) and \ref{Z300_Z800_contour}(b), respectively.  Note that for both cases, not only are the overall fusion energy yields significantly reduced, but also the optimal parameter space is significantly reduced.  In particular, the use of higher initial fuel densities has become very undesirable due to excessive bremsstrahlung losses.  These effects are important to keep in mind when considering future MagLIF designs, particularly when evaluating alternative (non-laser-based) preheating schemes.  We note, however, that some of these initial fuel density limitations can be circumvented by the use of a cryogenic DT ``ice" layer, such as that required by the ``high-gain" MagLIF concept of Ref.~\onlinecite{Slutz_PRL_2012}.

\begin{figure*}
\includegraphics[width=0.9\textwidth]{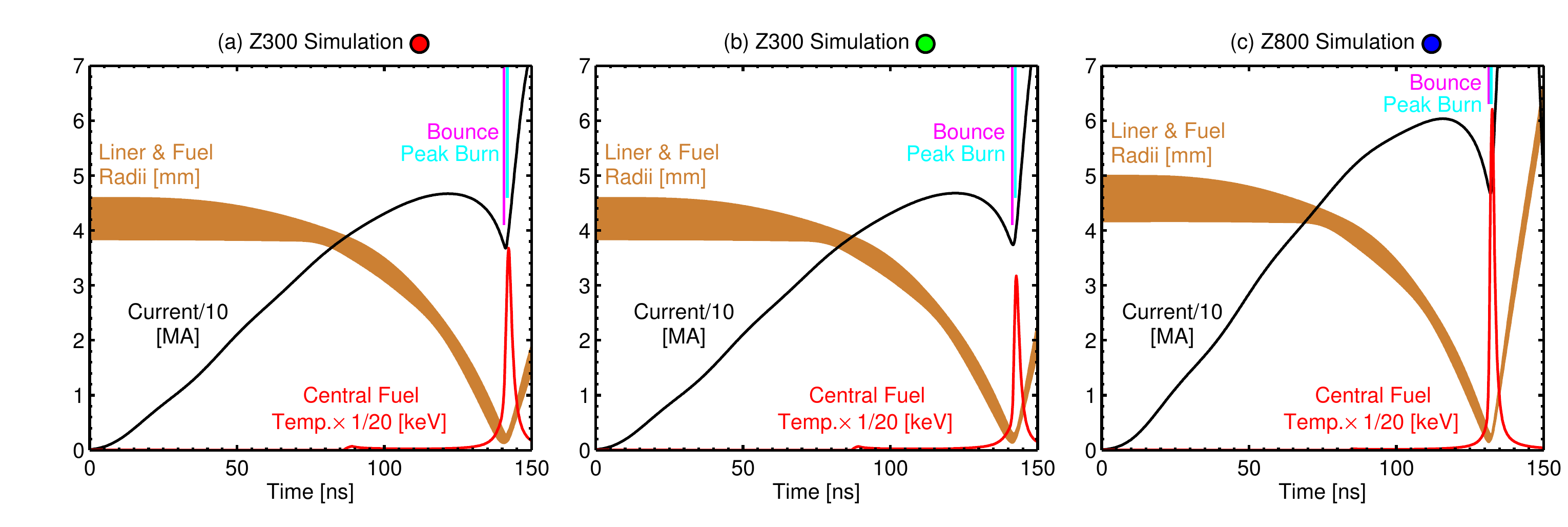}
\caption{\label{Z300_Z800_rt_montage}Implosion trajectories, drive currents, and central (on-axis) fuel temperatures for three SAMM simulations taken from the parameter scans of Fig.~\ref{future_contours}.  These cases correspond to the overlaid red, green, and blue circles in Figs.~\ref{future_contours}(b) and \ref{future_contours}(e).  All three cases are igniting examples where peak burn occurs after peak compression (as indicated by the vertical magenta and cyan timing marks).}
\end{figure*}

In Fig.~\ref{Z300_Z800_rt_montage}, we consider three specific cases taken from the parameter scans of Fig.~\ref{future_contours}.  These cases correspond to the overlaid circles in Figs.~\ref{future_contours}(b) and \ref{future_contours}(e).

In Fig.~\ref{Z300_Z800_rt_montage}(a), we present the results for a Z300 case with 9 kJ of preheat energy and an initial fuel density of 6 mg/cm$^3$.  Note that both the Z300 accelerator and a 9-kJ laser could fit in the footprint of today's Z and ZBL facilities.  In the simulation for this case, the target absorbs 4.3 MJ of the accelerator's energy by the time of peak fusion burn and returns a total fusion energy yield of 38 MJ.  Also, the maximum convergence ratio is 32 while the convergence ratio at peak burn is 26; however, in this igniting case, the peak burn occurs after maximum compression (during radial expansion) due to significant $\alpha$-heating rates.  The fact that this case ignites is also indicated by the strong explosion in the liner trajectory after maximum compression and by the central fuel temperature, which exceeds 70 keV. Additionally, about 5\% of the fuel mass is burned, $p\tau=57$~Gbar$\cdot$ns, and $p\tau_{({\text{no-}}\alpha)}=18$~Gbar$\cdot$ns.

In Fig.~\ref{Z300_Z800_rt_montage}(b), we present the results for Z300 with 20~kJ of preheat energy and an initial fuel density of 12 mg/cm$^3$.  Here, the target again absorbs 4.3 MJ of the accelerator's energy by the time of peak fusion burn, but this time the total fusion energy yield is 77 MJ.  Given that Z300's stored electrical energy is 48 MJ, this is a case where the overall machine gain exceeds unity.  Also, the maximum convergence ratio is 31 while the convergence ratio at peak burn is 23.  Due to ignition, peak burn again occurs after maximum compression.  In this case, the central temperature exceeds 60 keV at peak burn.  Additionally, about 5\% of the fuel mass is burned, $p\tau=62$~Gbar$\cdot$ns, and $p\tau_{({\text{no-}}\alpha)}=18$~Gbar$\cdot$ns.

In Fig.~\ref{Z300_Z800_rt_montage}(c), we present the results for Z800 with 30~kJ of preheat energy and an initial fuel density of 18 mg/cm$^3$.  Here, the target absorbs 8.2~MJ of the accelerator's energy by the time of peak fusion burn and returns a total fusion energy yield of 218 MJ.  Given that Z800's stored electrical energy is 130 MJ, this case represents a high yield scenario with an overall machine gain that exceeds unity.  Also, the maximum convergence ratio is 30 while the convergence ratio at peak burn is 23.  Due to ignition, peak burn again occurs after maximum compression (note the strong explosion in the liner trajectory after maximum compression).  In this case, the central temperature exceeds 120 keV at peak burn. Additionally, about 7\% of the fuel mass is burned, $p\tau=128$~Gbar$\cdot$ns, and $p\tau_{({\text{no-}}\alpha)}=28$~Gbar$\cdot$ns.

For all of the Z300 and Z800 results presented above, end losses were turned off in the simulations.  This is because, in the future, we hope to be able to design targets where the ends of the liner pinch off faster than the axial mid-plane of the liner to effectively (and cleanly) reduce the end losses.  This is a difficult challenge, and might not be possible without inadvertently mixing unwanted liner material into the fuel.  Thus, to test the effects of end losses, we reran the simulations for the three cases presented in Fig.~\ref{Z300_Z800_rt_montage} with single-ended end losses turned on and $r_{hole}=r_{ph0}$.  We found that the yields for the two Z300 cases were reduced by about 22\%, while the yield for the Z800 case was reduced by only 12\%.

Throughout this paper, we have considered only standard gas-burning MagLIF.  That is, we have not tried to account for the possibility of burning a cryogenic DT ice layer (this is presently beyond the scope of SAMM).  However, these cryogenic ``high-gain'' MagLIF scenarios have been studied in Ref.~\onlinecite{Slutz_PRL_2012}, where standard gas-burning MagLIF was compared to high-gain cryogenic MagLIF as a function of peak drive current out to about 70 MA (see Fig.~3 in Ref.~\onlinecite{Slutz_PRL_2012}).  The cases that we have presented here in Fig.~\ref{Z300_Z800_rt_montage} are in good agreement with the standard gas-burning yields presented in Fig.~3 of Ref.~\onlinecite{Slutz_PRL_2012}; however, we note that the ability to burn a cryogenic DT ice layer with a 60-MA driver could potentially result in another factor of 10 increase in fusion energy yield.

\begin{acknowledgments}
We would like to thank an anonymous referee for reviewing this manuscript and providing thoughtful insights and suggestions.  We would also like to thank the MagLIF/ICF and Dynamic Material Properties research groups in general for many useful technical discussions regarding various aspects of MagLIF. Finally, we thank the personnel of the Pulsed-Power Sciences Center (including the Z and ZBL facilities) at Sandia National Laboratories; without their hard work and dedication, this work would not have been possible.

Sandia is a multiprogram laboratory operated by Sandia Corporation, a Lockheed-Martin company, for the United States Department of Energy's National Nuclear Security Administration, under Contract No. DE-AC04-94AL85000.
\end{acknowledgments}


\begin{thebibliography}{61}
\bibitem [{las({\natexlab{a}})}]{laser_improvement_note}%
  \BibitemOpen
  \href@noop {} {}\bibinfo {howpublished} {Additional laser-only studies are
  presently underway to better understand and improve the laser transmission
  and coupling to the fuel. These activities have been taking place at both
  Sandia and the Laboratory for Laser Energetics at the University of
  Rochester. Also, more recently, design efforts have commenced for preheat
  experiments on the National Ignition Facility at Lawrence Livermore National
  Laboratory.}\BibitemShut {Stop}%
\bibitem [{lec()}]{lecapproxc}%
  \BibitemOpen
  \href@noop {} {}\bibinfo {howpublished} {In reality, the laser energy that is
  deposited in the fuel of the laser entrance channel can still couple to the
  fuel of the imploding region through various axial transport mechanisms like
  shockwaves and thermal conduction. However, SAMM presently does not account
  for this possibility. Thus, the values reported here for the amount of laser
  energy that could have gotten past the laser entrance window should be
  considered first-order approximations.}\BibitemShut {Stop}%
\end{thebibliography}
%

\end{document}